\documentclass[10pt,superscriptaddress,twocolumn,nofootinbib]{revtex4}

\input epsf
\usepackage{amsmath}
\usepackage{amssymb}
\usepackage{amsthm}

\begin{document}

\title{Quantum computing and polynomial equations over the finite
field $\mathbb{Z}_2$}

\author{Christopher~M.~Dawson} \email{dawson@physics.uq.edu.au}
\affiliation{School of Physical Sciences, The University of
Queensland, Brisbane, Queensland 4072, Australia}
\affiliation{Centre for Quantum Computer Technology, The
University of Queensland, Brisbane, Queensland 4072, Australia}

\author{Henry~L.~Haselgrove} \email{HLH@physics.uq.edu.au}
\affiliation{School of Physical Sciences, The University of
Queensland, Brisbane, Queensland 4072, Australia}
\affiliation{Information Sciences Laboratory, Defence Science and
Technology Organisation, Edinburgh 5111, Australia}

\author{Andrew~P.~Hines} \email{hines@physics.uq.edu.au}
\affiliation{School of Physical Sciences, The University of
Queensland, Brisbane, Queensland 4072, Australia}
\affiliation{Centre for Quantum Computer Technology, The
University of Queensland, Brisbane, Queensland 4072, Australia}

\author{Duncan~Mortimer} \email{duncanm@physics.uq.edu.au}
\affiliation{School of Physical Sciences, The University of
Queensland, Brisbane, Queensland 4072, Australia}
\affiliation{Centre for Quantum Computer Technology, The
University of Queensland, Brisbane, Queensland 4072, Australia}

\author{Michael~A.~Nielsen} \email{nielsen@physics.uq.edu.au}
\homepage{www.qinfo.org/people/nielsen} \affiliation{School of
Physical Sciences, The University of Queensland, Brisbane,
Queensland 4072, Australia} \affiliation{School of Information
Technology and Electrical Engineering, The University of
Queensland, Brisbane, Queensland 4072, Australia}

\author{Tobias~J.~Osborne} \email{T.J.Osborne@bristol.ac.uk}
\affiliation{School of Mathematics, University of Bristol,
University Walk, Bristol BS8 1TW, United Kingdom}

\date{\today}

\begin{abstract}
  What is the computational power of a quantum computer?  We show that
  determining the output of a quantum computation is equivalent to
  counting the number of solutions to an easily computed set of
  polynomials defined over the finite field $\mathbb{Z}_2$. This
  connection allows simple proofs to be given for two known relationships
  between quantum and classical complexity classes, namely
  $\mathbf{BQP}\subseteq\mathbf{P}^{\mathbf{\#P}}$ and $\mathbf{BQP} \subseteq
  \mathbf{PP}$.
\end{abstract}

\maketitle

\section{Introduction}

%
% intro: quantum computers
%
Quantum computers have stimulated great interest due to their promise
of being able to solve problems considered infeasible on conventional
classical computers~\cite{Shor97a,Shor94a}.  This interest has led to
rapid developments in physics, mathematics, and computer
science~\cite{Nielsen00a,Preskill98c}.  One of the central open
problems in quantum computation is to precisely characterize the power
of quantum computers, i.e., what problems they can and cannot solve
efficiently.

%
% what we do in this paper
%
In this paper we show that determining the output of a quantum
computation is equivalent to counting the number of solutions to
certain sets of polynomial equations over the finite field
$\mathbb{Z}_2$.  Equivalently, in the language of algebraic geometry,
this means counting the number of points in an algebraic variety.  The
proof combines Feynman's sum-over-paths formulation of quantum
mechanics~\cite{Feynman49a,Feynman65e} with a description of quantum
computing in terms of a universal set of quantum gates specially
chosen to make the sum-over-paths take a simple form.

This reformulation of quantum computation is interesting for
several reasons. First, it reveals a connection between quantum
computation and one of the central problems in algebraic geometry.
Indeed, much of the development of modern algebraic
geometry~\cite{Cox97a,Hartshorne77a} has been driven by the
problem of counting the points in an algebraic variety, e.g., this
problem gave rise to the well-known Weil
conjectures~\cite{Weil49a}.  Second, it reveals a connection
between quantum computation and computational complexity. In
particular, computational complexity theorists have shown that the
problem of counting solutions to polynomials over finite fields is
$\mathbf{\#P}$-complete ~\cite{Ehrenfeucht90a,Gathen97a}.

As a consequence of this connection, our result has as a corollary a
simple proof of one of the sharpest known results relating quantum and
classical complexity classes\footnote{For a general overview of
  computational complexity theory, see~\cite{Papadimitriou94a}.  For
  definitions and references on all the complexity classes we consider
  here, and many others, an excellent reference is Aaronson's
  ``Complexity Zoo''~\cite{AaronsonZoo}.}~\cite{Bernstein97a},
$\mathbf{BQP}\subseteq\mathbf{P}^{\mathbf{\#P}}$, where $\mathbf{BQP}$
is, informally, the class of decision problems efficiently soluble on
a quantum computer.  The complexity class $\mathbf{P}^{\mathbf{\#P}}$
is, in turn, a subset of the well-known complexity class
$\mathbf{PSPACE}$ of problems requiring polynomial space (but possible
exponential time) to solve on a classical computer.  Proving
$\mathbf{P} \neq \mathbf{PSPACE}$ would represent a major breakthrough
in classical complexity theory.  A consequence of the result
$\mathbf{BQP} \subseteq \mathbf{PSPACE}$ is that any proof that
quantum computers are more efficient than classical computers will
imply $\mathbf{P} \neq \mathbf{PSPACE}$, and thus would have major
implications for classical computational complexity.

Our techniques also imply a simple proof of a result even sharper
than $\mathbf{BQP} \subseteq \mathbf{P^{\#P}}$, namely
$\mathbf{BQP} \subseteq \mathbf{PP}$~\cite{Adleman97a}.  To our
knowledge, this is the sharpest known relation between
$\mathbf{BQP}$ and a natural classical complexity class\footnote{A
stronger relation, $\mathbf{BQP} \subseteq
  \mathbf{AWPP}$ has been proved by Fortnow and
  Rogers~\cite{Fortnow99a}.  However, as they note, $\mathbf{AWPP}$ is
  a rather artifical complexity class.}.

%
% compare our techniques with those used by others
%
As described above, our approach is based on the sum-over-paths
formulation of quantum mechanics.  Interestingly, the papers just
mentioned~\cite{Bernstein97a,Adleman97a,Fortnow99a} all use variants
of the sum-over-paths formulation to obtain their relations between
$\mathbf{BQP}$ and various classical complexity classes.  This is also
true of the paper by Knill and Laflamme~\cite{Knill01a}, which
connects quantum computation to the problem of estimating quadratic
weight enumerators.  What all these papers share in common is that in
evaluating the amplitude for a particular path through a quantum
computation, it is necessary to keep track of both the \emph{phase} of
the amplitude, and also of the \emph{magnitude} of the amplitude.  By
choosing a particular universal gate set, we simplify the problem so
that it is necessary to keep track only of the phase, not of the
magnitude.  By doing this, the relationship between quantum computing
and polynomial equations arises in a natural way.

%
% outline
%
The structure of the paper is as follows.  We begin our account in
Section~\ref{sec:polygen}, with an example showing how the matrix
elements of a unitary quantum circuit may be computed by counting the
number of solutions to a set of polynomial equations over
$\mathbb{Z}_2$.  This example then motivates a general argument, given
in Section~\ref{sec:genproof}, showing that the matrix elements of a
quantum circuit may always be computed by counting the solutions to an
appropriate set of polynomial equations.
Section~\ref{sec:applications} discusses the implications of this
result for the solution of decision problems on quantum computers, and
for the relation between quantum and classical complexity classes.
Section~\ref{sec:sampling} discusses whether or not it is possible to
combine our approach with Monte Carlo sampling techniques to obtain an
efficient classical procedure for simulating quantum computers;
unsurprisingly, this attempt fails, but the failure is instructive.
Section~\ref{sec:other-gate-sets} describes various possible
reformulations of our results, and some prospects for further
simpilification.  We conclude in Section~\ref{sec:conclusion}.

%
% The example
%
\section{Example of how to find the polynomials}
\label{sec:polygen}

\begin{figure}%[!tbhp]
\begin{center}
\epsfxsize=5cm \epsfbox{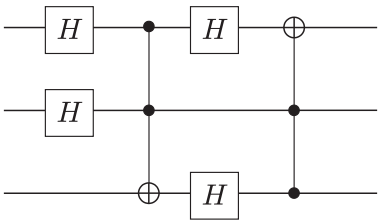}
 \caption{An example $N=3$ quantum circuit containing two Toffoli gates and four Hadamard gates.}
 \label{fig:example_circuit}
\end{center}
\end{figure}

In this section, we provide an example showing how to calculate a
transition amplitude for a quantum circuit by counting the number of
solutions to sets of polynomial equations. We explain the method with
the aid of a simple example quantum circuit, but delay the general
proof until the next section.

%
% beginning of the example: introduce the universal gate set
%
All our results rely on constructing circuits out of certain specially
chosen sets of universal gates.  For our initial discussion we will
use the Toffoli and Hadamard gates, which have been shown to be
universal for quantum computation by Shi~\cite{Shi03a} and
Aharonov~\cite{Aharonov03c}\footnote{Note that the Hadamard and
  Toffoli gates have only real matrix elements, so cannot generate a
  dense subset of the full unitary group. The universality proofs
  of~\cite{Shi03a,Aharonov03c} use simple encodings to achieve
  universality.}.  Later we'll discuss the general properties of a
gate set necessary to make our style of argument work.  Recall that a
Toffoli gate has action on computational basis states given by
$|x,y,z\rangle \rightarrow |x,y,z\oplus xy\rangle$.  The single-qubit
Hadamard gate maps $|0\rangle$ to $(|0\rangle+|1\rangle)/\sqrt 2$ and
$|1\rangle$ to $(|0\rangle-|1\rangle)/\sqrt 2$.

Suppose we are given some $N$-qubit quantum circuit, constructed from
Toffoli and Hadamard gates.  A simple example for $N = 3$ is shown in
Fig.~\ref{fig:example_circuit}. Imagine, further, that we wish to
calculate the matrix element $\langle \mathbf{b} | U | \mathbf{a}
\rangle$, where $U$ is the unitary action of the quantum circuit, and
where $|\mathbf{a}\rangle$ and $|\mathbf{b}\rangle$ are {\em
  computational basis states} (that is, $|\mathbf{a}\rangle=|a_0,a_1,
\cdots,a_{N-1}\rangle$, where $(a_0, a_1, \ldots, a_{N-1})\in
\mathbb{Z}_2^{\times N}$ is a bit string, and similarly for
$|\mathbf{b}\rangle$).

To determine $\langle \mathbf{b}|U|\mathbf{a}\rangle$ we will
define the notion of a set of \emph{allowed} or \emph{admissible}
classical paths through the circuit, from input $(a_0, a_1,
\ldots, a_{N-1})$ to output $(b_0, b_1, \ldots, b_{N-1})$, and the
corresponding {\em
  phases} associated with each of those paths. To define the allowed
classical paths, we first define what we shall call a \emph{classical}
version of the quantum circuit in Fig.~\ref{fig:example_circuit}.
This is a \emph{classical} circuit which is formed by replacing each
of the qubits in Fig.~\ref{fig:example_circuit} with a classical bit,
by replacing the quantum Toffoli gate by a classical Toffoli gate,
which takes $(a_1, a_2, a_3)$ to $(a_1, a_2, a_3\oplus a_1a_2)$, and
by replacing each Hadamard gate by what we shall call a
\emph{classical Hadamard gate}, which, regardless of its input, may
output \emph{either} a zero or a one.  To describe this situation we
introduce a \emph{path variable} $x\in\mathbb{Z}_2$ to denote the
output of the Hadamard:
\begin{equation}
\epsfxsize=2.3cm \epsfbox{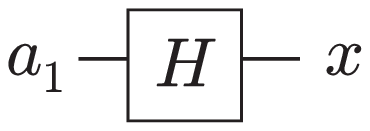}.
\end{equation}
A classical path is then a sequence of classical bit strings,
$\mathbf{a}, \mathbf{a'},\mathbf{a''},\ldots$, with the respective
bit strings corresponding to the state of the classical circuit
after each gate has been applied.  A classical path is said to be
\emph{allowed} or \emph{admissible} if there exists a choice of
the path variables giving rise to that path.  If there are $h$
Hadamard gates in a particular circuit, then there will be $2^h$
distinct allowed classical paths, with each path corresponding to
a different choice of the path variables $x_1,x_2,\ldots,x_h$.

\begin{figure}%[!tbhp]
\begin{center}
\epsfxsize=7cm \epsfbox{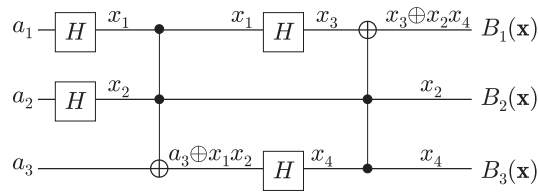}
 \caption{The wires in the example circuit have been annotated with
 the allowed classical paths starting with input $(a_1,a_2,a_3)$. }
 \label{fig:example}
\end{center}
\end{figure}

In Fig.~\ref{fig:example}, these rules have been applied to construct
a classical circuit from the example quantum circuit.
Fig.~\ref{fig:example} shows all allowed classical paths from the
input string $(a_1,a_2,a_3)$, as a function of the path variables
$(x_1,\dots,x_4)\equiv\mathbf{x}$.  Note that there are $2^4=16$
admissible paths corresponding to all the possible bit assignments of
the variables $\mathbf{x}$. The output bit values, denoted
$B_j(\mathbf{x})$, are polynomial functions of the $x_j$, i.e.,
elements of the polynomial ring $\mathbf{Z}_2[\mathbf{x}]$. In this
example, $B_1(\mathbf{x})=x_3\oplus x_2x_4$, $B_2(\mathbf{x})=x_2$,
and $B_3(\mathbf{x})=x_4$.

We now define the \emph{phase} $\phi(\mathbf{x})$ of an allowed
classical path with path variables $\mathbf{x}$. This definition may
appear a little mysterious at first; its importance will become
clearer below.  We define the phase to be the sum (modulo $2$) over
all (classical) Hadamard gates, of the product of the input and output
bit values of the gate:
\begin{equation}
\phi(\mathbf{x})\equiv
% \bigoplus ?
\sum_{\mbox{\scriptsize Hadamard gates}}(\mbox{input
value})(\mbox{output value}).
\end{equation}
$\phi(\mathbf{x})$ is a polynomial function of the $x_j$. In this
example $\phi(\mathbf{x})=a_1x_1\oplus a_2x_2 \oplus x_1x_3 \oplus
x_4(a_3 \oplus x_1x_2)$.

It turns out that the matrix element
$\langle\mathbf{b}|U|\mathbf{a}\rangle$ is given by the following sum
over allowed paths from $\mathbf{a}$ to $\mathbf{b}$:
\begin{equation}
\langle\mathbf{b}|U|\mathbf{a}\rangle= \frac{1}{\sqrt{2^h}}
\sum_{\mathbf{x}:\hspace{.5ex}\mathbf{B}(\mathbf{x})=\mathbf{b}}
(-1)^{\phi(\mathbf{x})}. \label{eq:sum}
\end{equation}
We will prove this explicitly in the next section; however, we hope
this expression is at least plausible to the reader, expressing the
transition amplitude as a sum over the allowed paths through the
circuit, with amplitudes of the appropriate magnitude and phase.  The
terms in the sum all have the same absolute value, but vary in sign.
We define $\#(0)$ and $\#(1)$ to be the number of positive and
negative terms in the sum respectively. That is,
\begin{equation}
\#(0)=|\, \{ \mathbf{x} \,|\, \mathbf{B}(\mathbf{x}) = \mathbf{b} \mbox{
and } \phi(\mathbf{x})=0 \}\,|, \label{eq:num0}
\end{equation}
and
\begin{equation}
\#(1)=| \, \{ \mathbf{x} \,|\, \mathbf{B}(\mathbf{x}) = \mathbf{b} \mbox{
and } \phi(\mathbf{x})=1 \}\,|. \label{eq:num1}
\end{equation}
Eq.~(\ref{eq:sum}) can now be written as
\begin{equation}
\langle\mathbf{b}|U|\mathbf{a}\rangle = \frac{1}{\sqrt{2^h}} [
\#(0) - \#(1)]. \label{eq:amplitude}
\end{equation}
The expressions in Eqs.~(\ref{eq:num0}) and (\ref{eq:num1}) each count
solutions to a system of $N+1$ polynomials in $h$ variables over the
field $\mathbb{Z}_2$. We can make some general remarks about the
properties of these polynomials. If we ensure that each Toffoli gate
is followed by a Hadamard gate on the target line (if necessary by
inserting pairs of Hadamard gates, which act as the identity gate),
then the polynomials $B_j(\mathbf{x})$ will have at most two terms and
have order at most two, and $\phi(\mathbf{x})$ will have at most $2h$
terms and order at most three.

As a simple example of this method in action, let's calculate
$\langle\mathbf{0}|U|\mathbf{0}\rangle$ in the example circuit.
$\mathbf{B}(\mathbf{x})=\mathbf{0}$ has two solutions, $x_1=0$ or $1$,
$x_2=0$, $x_3=0$, $x_4=0$. For each solution $\phi(\mathbf{x})=0$, so
$\#(0)=2$ and $\#(1)=0$. Thus, $\langle \mathbf{0} | U |
\mathbf{0}\rangle = 2/\sqrt{16}=1/2$, which is easily verified to be
the correct amplitude.

%
% A proof that the recipe works
%

\section{General proof}
\label{sec:genproof}

In this section we prove that the method in the previous section works
for any quantum circuit made from Hadamard and Toffoli gates.  To do
this, we express the unitary action $U$ of the circuit as a product of
gates:
\begin{equation}\label{eq:u}
U=U^{(M)}U^{(M-1)}\dots U^{(2)}U^{(1)}.
\end{equation}
Each of the gates $U^{(m)}$ represents either a Hadamard acting on
one qubit, or a Toffoli gate acting on three qubits. Each
$U^{(m)}$ will thus act as an identity on all but one or three of
the qubits.

For the general case in Eq.~(\ref{eq:u}), it is clear that we can
write
\begin{equation}
\begin{split}
\langle \mathbf{b} |U| \mathbf{a} \rangle =
 \sum_{\mathbf{c},\mathbf{d},\dots,\mathbf{z}} \langle \mathbf{b} | &
U^{(M)} |\mathbf{z}\rangle \langle \mathbf{z} | U^{(M-1)} | \mathbf{y}
\rangle \, \dots \\
 & \, \, \dots \, \langle \mathbf{d} | U^{(2)} | \mathbf{c} \rangle
\langle \mathbf{c} | U^{(1)} | \mathbf{a} \rangle
\end{split}
\label{eq:mult}
\end{equation}
where $\mathbf{c},\dots,\mathbf{z}$ are each bit strings of length
$N$. Eq.~(\ref{eq:mult}) expresses the transition amplitude of the
entire circuit from the initial state $\mathbf{a}$ to the final state
$\mathbf{b}$ as a sum of amplitudes over all sequences of
computational states $\mathbf{a}\rightarrow \mathbf{c} \rightarrow
\mathbf{d} \rightarrow$ \dots $ \rightarrow \mathbf{z} \rightarrow
\mathbf{b}$ through the circuit. For each of these sequences, the
corresponding term in the sum is given by a product of contributions
from each of the gates $U^{(m)}$. This is already very close to the
``sum over classical paths'' approach used in the previous section. To
complete the connection, we must first show that the sum in
Eq.~(\ref{eq:mult}) may be restricted to the set of allowed classical
paths as defined in the previous section, and second, that those
remaining terms in the sum are equal to the terms in
Eq.~(\ref{eq:sum}).

Consider a factor $\langle\mathbf{s}|U^{(m)}|\mathbf{r}\rangle$ from
Eq.~(\ref{eq:mult}). Say that $U^{(m)}$ acts as the identity on qubit
$k$. Then, if $s_k\neq r_k$, the factor
$\langle\mathbf{s}|U^{(m)}|\mathbf{r}\rangle$ will be zero.  Suppose
that this is not the case for any of the qubits on which $U^{(m)}$
acts as the identity, and that $U^{(m)}$ is a Toffoli gate acting on
qubits $k_1$, $k_2$ and $k_3$. From the definition of the Toffoli
gate, $\langle\mathbf{s}|U^{(m)}|\mathbf{r}\rangle$ will equal one if
$s_{k_1}=r_{k_1}$, $s_{k_2}=r_{k_2}$, and $s_{k_3}=r_{k_3}\oplus
r_{k_1}r_{k_2}$, and will equal zero otherwise.  Alternatively,
suppose $U^{(m)}$ is a Hadamard gate acting on qubit $k$. Then, from
the definition of the gate,
$\langle\mathbf{s}|U^{(m)}|\mathbf{r}\rangle$ will equal $-1/\sqrt{2}$
if $s_k=r_k=1$ (that is, if $s_kr_k=1$), and will equal $1/\sqrt{2}$
otherwise.

%Thus,
%the terms in Eq.~(\ref{eq:mult}) can only be nonzero when the
%sequence $\mathbf{a}\rightarrow \mathbf{c} \rightarrow$ \dots $
%\rightarrow \mathbf{z} \rightarrow \mathbf{b}$ corresponds to a
%{\em classical path} as defined in the previous section, that is
%if the sequence gives only one bit value for each uninterrupted
%horizontal wire segment.

From these observations, we see that the terms in
Eq.~(\ref{eq:mult}) are nonzero only when $\mathbf{a}\rightarrow
\mathbf{c} \rightarrow$ \dots $ \rightarrow \mathbf{z} \rightarrow
\mathbf{b}$ is an {\em
  allowed classical path}.  Furthermore, the absolute value of any
nonzero term will be equal to $1/\sqrt{2^h}$, where $h$ is the number
of the gates $U^{(m)}$ that are Hadamards. Further, the sign of such a
term will be $(-1)^p$, where $p$ is defined by summing over the
product of the input and output values to each classical Hadamard
gate.  Thus, we have proven Eq.~(\ref{eq:sum}) and hence
Eq.~(\ref{eq:amplitude}), as required.

\section{Application to decision problems and computational complexity}
\label{sec:applications}

%
% applications
%
The results described in the previous section have interesting
consequences in the special case when the quantum computer is being
used to solve a decision problem, i.e., a problem where the answer is
either ``yes'' or ``no''.  Many interesting problems are either
explicitly decision problems, e.g., satisfiability, or can be recast
as equivalent decision problems~\cite{Papadimitriou94a}, e.g.,
factoring and the travelling salesman problem.

\emph{A priori} it is not obvious that the ability to calculate, even
extremely efficiently, the matrix elements of quantum circuits is
particularly useful. This is because there are exponentially many
matrix elements for a given quantum circuit.  It would appear that
even a constant-time recipe to calculate matrix elements would, at
best, provide an exponential method to estimate the output of a given
quantum circuit. However, following~\cite{Bennett97a}, we now show
that when a quantum circuit is being used to solve a decision problem,
the output of the quantum circuit can be inferred from knowledge of
one fixed matrix element.

\begin{figure}%[!tbhp]
\begin{center}
\epsfxsize=7cm \epsfbox{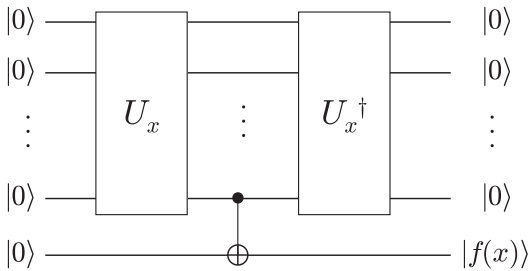}
 \caption{Reformulating a quantum circuit for a decision problem so that it only outputs states of the form $|\mathbf{0}\rangle|0\rangle$ or $|\mathbf{0}\rangle|1\rangle$. }
 \label{fig:decision}
\end{center}
\end{figure}

Consider a quantum circuit $U_x$ for an instance $x$ of an arbitrary
decision problem. We suppose for now that the quantum circuit is
deterministic, i.e., it gives the correct output with probability $1$.
We may also suppose, without loss of generality, that the output
(``no'' or ``yes'') is indicated by the value of the first output
qubit $Q_1$, as $|0\rangle$ or $|1\rangle$, respectively. The
remaining output qubits $Q_2\cdots Q_N$ are left in a ``junk'' quantum
state $|\phi\rangle$.  By adjoining an ancilla qubit $Q_A$ initialised
in the state $|0\rangle$ and then applying a {\sc cnot} gate on $Q_1$
and $Q_A$ the answer $f(x)$ can be copied into $Q_A$ (see
Fig.~\ref{fig:decision}). Finally, applying the inverse operation
$U_x^\dag$ to qubits $Q_1\cdots Q_N$ \emph{uncomputes} the output
state to the initial state $|\mathbf{0}\rangle$. The state of the
quantum computer is now $|\mathbf{0}\rangle_{Q_{1}\cdots
  Q_{N}}|f(x)\rangle_{Q_A}$. Without loss of generality it is
therefore possible to assume that \emph{any} quantum circuit for an
instance $x$ of a decision problem outputs one of only \emph{two}
possible states: $|\mathbf{0}\rangle|0\rangle$ or
$|\mathbf{0}\rangle|1\rangle$.  It follows that knowledge of only
\emph{one} matrix element is enough to understand the output of such a
quantum circuit.

A useful corollary of this construction is that to determine the
output of the quantum circuit it is actually sufficient to determine
the \emph{sign} of a single amplitude.  To see this, suppose in the
construction above that we had applied a {\sc not} gate and a Hadamard
first to the ancilla qubit, so it was in the state
$(|0\rangle-|1\rangle)/\sqrt 2$.  Applying $U_x$, the {\sc cnot} and
$U_x^\dagger$, as was done in the earlier construction, followed by
another Hadamard to the ancilla qubit, gives as output
$(-1)^{f(x)}|\mathbf{0}\rangle|0\rangle$.  It follows that knowledge
of just the sign of the amplitude determines $f(x)$. In consequence of
this observation, through the sequel we will change our notation
somewhat, and denote the relevant matrix element as $\langle
\mathbf{0}| U_x |\mathbf{0}\rangle$.

In making this argument, we have assumed that the quantum circuit
solves the decision problem with probability $1$, i.e.,
deterministically.  However, following the lines of~\cite{Bennett97b},
a more elaborate calculation based on the same idea shows that a
similar result holds even for a quantum circuit which only outputs the
correct answer probabilistically, provided the probability is bounded
below by some appropriate constant, say $3/4$.  Thus, to determine the
output of a quantum circuit which solves some decision problem it
suffices to determine the sign of a single matrix element $\langle
\mathbf{0}| U_x |\mathbf{0}\rangle$.

%
% quantum computational complexity
%
In the context of quantum computational complexity our results provide
as an immediate corollary the result $\mathbf{BQP} \subseteq
\mathbf{P}^{\mathbf{\#P}}$, due to Bernstein and
Vazirani~\cite{Bernstein97a}.  We now outline the proof.  Recall that
the class $\mathbf{P}^{\mathbf{\#P}}$ is defined to be the class of
decision problems decidable in polynomial time with the aid of an
oracle which computes, at unit cost, the solution to a problem
complete for $\mathbf{\#P}$.  To see that $\mathbf{BQP} \subseteq
\mathbf{P}^{\mathbf{\#P}}$, we suppose our oracle $O(\mathbf{p})$
accepts as input a set of polynomials
$\mathbf{p}\in\mathbb{Z}_2[\mathbf{x}]$, and returns the number of
solutions of the corresponding set of equations over $\mathbb{Z}_2$.
Supplied with such an oracle it is clear from the results of
Sections~\ref{sec:polygen} and~\ref{sec:genproof} that we can
determine the amplitudes at the output of a uniformly generated
quantum circuit of polynomial size, and thus decide any language in
$\mathbf{BQP}$.  Because the problem of counting solutions to
polynomial equations over $\mathbb{Z}_2$ is clearly in $\mathbf{\#P}$,
we obtain the inclusion
$\mathbf{BQP}\subseteq\mathbf{P}^{\mathbf{\#P}}$.

Our techniques can also be used to prove the stronger inclusion
$\mathbf{BQP} \subseteq \mathbf{PP}$, due to Adleman, Demarrais, and
Huang~\cite{Adleman97a}.  To see this, recall the definition of
$\mathbf{PP}$ due to Fenner, Fortnow and
Kurtz~\cite{Fenner94a}\footnote{The original definition is due
  to~\cite{Gill77a}, but is more unwieldly in the present context.}.
First, define a $\mathbf{GapP}$ function to be a function expressible
as the difference of two $\mathbf{\#P}$ functions.  Then the class
$\mathbf{PP}$ consists of all those languages $L$ such that for some
$\mathbf{GapP}$ function $f$, and for all $x$, either (a) if $x$ is in
$L$, then $f(x) > 0$; or (b) if $x$ is not in $L$ then $f(x) < 0$.

The proof that $\mathbf{BQP}$ is a subset of $\mathbf{PP}$ is now
trivial.  The transition amplitude $\langle \mathbf{0}|U_x
|\mathbf{0}\rangle$ may be written as
\begin{eqnarray}
  \langle \mathbf{0}|U_x |\mathbf{0}\rangle = \frac{\#(0)-\#(1)}{\sqrt{2^h}},
\end{eqnarray}
which makes it a difference of two $\mathbf{\# P}$ functions, $\#(0)$
and $\#(1)$, and thus the amplitude $\langle \mathbf{0}|U_x
|\mathbf{0}\rangle$ is a $\mathbf{GapP}$ function.  The result follows
using~\cite{Fenner94a}'s definition of $\mathbf{PP}$.

\section{Sampling methods for simulating quantum circuits}
\label{sec:sampling}

An interesting question to ask in the light of our results is whether
they can provide a means of speeding up the simulation of quantum
computers by classical means.  An obvious technique for doing this is
to use Monte Carlo sampling to estimate the number of solutions to the
equations of interest.  Unfortunately, this technique fails to work in
general.

%In order to see why standard Monte Carlo sampling fails, consider the
%following general description of a sampling algorithm.
%We have a
%finite universe $U$ of known size $|U|$. The goal is to estimate the
%size of some set $G\subset U$ of an unknown size. In our case $U$ is
%the set of all admissible paths through the circuit. A trial for
%estimating $|G|$ consists of randomly and uniformly select $s\in U$,
%and then test to see if $s\in G$.  An elementary result from
%probability theory, Bernstein's inequality (see, e.g., page~31
%of~\cite{Grimmett92a}) shows that if we perform $N=4\ln(2/\delta)
%/\epsilon^2$ trials, then the number $Y$ of trials where an element of
%$G$ is chosen satisfies
%\begin{equation}
%\text{Pr}[|Y-|G| | \leq N \epsilon] \ge 1-\delta.
%\end{equation}
%This can be used to estimate the size of the set $|G|$.

To see why sampling fails, let's look at how such an algorithm might
work.  We suppose we have worked out the polynomials $B_j(x_1, x_2,
\ldots, x_{h})$, where $j$ ranges over the $N$ output qubits, and
there are $h$ Hadamard gates.  For a decision problem, an admissible
path $\mathbf{x}$ for the quantum circuit is one for which
$B_j(\mathbf{x})=0$, $\forall j=1, \ldots, N$, corresponding to
calculating the matrix element
$\langle\mathbf{0}|U_x|\mathbf{0}\rangle$.  As before we define
\begin{equation}
\#(0)=|\, \{ \mathbf{x} \,|\, \mathbf{B}(\mathbf{x}) = \mathbf{0} \mbox{
and } \phi(\mathbf{x})=0 \}\,|,
\end{equation}
and
\begin{equation}
\#(1)=| \, \{ \mathbf{x} \,|\, \mathbf{B}(\mathbf{x}) = \mathbf{0} \mbox{
and } \phi(\mathbf{x})=1 \}\,|,
\end{equation}
so that
\begin{equation}
\langle\mathbf{0}|U|\mathbf{0}\rangle = \frac{1}{\sqrt{2^h}} [
\#(0) - \#(1)].
\end{equation}
To estimate $\langle \mathbf{0}|U|\mathbf{0}\rangle$ we need to
estimate both $\#(0)$ and $\#(1)$.  \emph{A priori} it is possible
that $\#(0)$ and $\#(1)$ scale as $2^h$, so to obtain an estimate of
$\langle\mathbf{0}|U|\mathbf{0}\rangle$ accurate to some constant
precision requires a number of trials exponential in $h$.  It follows
from these observations that Monte Carlo sampling is not, in general,
an efficient way of estimating amplitudes for quantum computation.

%We also encounter a second major problem. We cannot provide any
%\emph{a priori} bounds on the fraction $|U|/|G|$ --- we cannot
%even calculate $|U|$ efficiently! It is plausible that this
%fraction could be very large (especially in a ``yes'' instance).
%To get sufficient confidence in the output of the approximation
%algorithm we need to make, in the worst case, exponentially many
%(in $h$) trials.

\section{Other gate sets}
\label{sec:other-gate-sets}

What other universal gate sets might be amenable to the path-sums
approach? Might it be possible to find universal sets giving rise to
sets of polynomial equations for which it is possible to efficiently
estimate the number of solutions?  To answer these questions, note
that the key property of the gate set used in the path-sum approach is
that the amplitudes in each gate are \emph{balanced}, i.e., the
non-zero elements of the gate all have the same absolute value.  For
example, the non-zero elements of the Hadamard gate all have absolute
value $1/ \sqrt{2}$, while the non-zero elements of the Toffoli gate
all have absolute value $1$.  Given any universal set of balanced
gates, it is not difficult to write down a sum-over-paths formulation
along similar lines to that done for the Hadamard-Toffoli set in
Sections~\ref{sec:polygen} and~\ref{sec:genproof}.

Rather than a general discussion, we will simply give a single
informative example of the results obtained when this approach is
followed, basing our discussion on the universal set consisting of
$T=\left(\begin{smallmatrix} 1&0 \\ 0& e^{\frac{\pi
        i}{4}}\end{smallmatrix}\right)$, $H$, and {\sc cnot}.
Applying the procedure described in Sections~\ref{sec:polygen}
and~\ref{sec:genproof} to a quantum circuit composed only of $T$, $H$,
and {\sc cnot} shows that that an amplitude $\langle \mathbf{b} |U|
\mathbf{a}\rangle$ can be found according to the following recipe.
First, define a classical circuit corresponding to the quantum circuit
by replacing all qubits by bits, quantum {\sc cnot}s by classical {\sc
  cnot}s, $T$ gates by identity operations, and Hadamard gates by
classical Hadamard gates, as earlier.  Notions of path variables and
admissible paths are defined as before; note that the polynomials
$B_j(\mathbf{x})$ are now \emph{linear} in the path variables
$\mathbf{x}$, due to the linearity of the gates appearing in the
classical circuit.  We can thus write the transition amplitude
\begin{equation}\label{eq:pathsum}
\langle \mathbf{b} |U|\mathbf{a}\rangle = \sum_{\mathbf{x}: B_j(\mathbf{x}) =
\mathbf{b}}\Phi(\mathbf{x}),
\end{equation}
where the phase factor $\Phi(\mathbf{x})$ depends on the value of the
path variables $\mathbf{x}$.  Following an argument analogous to that
in Sections~\ref{sec:polygen} and~\ref{sec:genproof}, we find that
these phases can be written in the form $\exp(i \pi
\varphi(\mathbf{x})/4)$, where $\varphi(\mathbf{x})$ is a polynomial
in the path variables $\mathbf{x}$; it is a polynomial in mixed
arithmetic involving multiplication over $\mathbb{Z}_2$ and addition
over $\mathbb{Z}_8$.  Because the $B_j$ are linear, it is possible to
eliminate variables, moving to a new set of unconstrained variables
$\mathbf{y} \in \mathbb{Z}_2^{\times h'}$, and writing the transition
amplitude as
\begin{equation}
\langle \mathbf{b} |U|\mathbf{a}\rangle = \sum_{\mathbf{y}}
 \exp(i \pi \phi(\mathbf{y})/ 4),
\end{equation}
where $\phi(\mathbf{y})$ is again a polynomial in mixed arithmetic
involving multiplication over $\mathbb{Z}_2$ and addition over
$\mathbb{Z}_8$; with a little work, it can easily be verified that
$\phi$ is of order at most two.  It follows that all we need do is
count the number of solutions to the eight equations $\phi(\mathbf{y})
= 0, \phi(\mathbf{y}) = 1, \ldots, \phi(\mathbf{y}) = 7$, in order to
determine such a transition amplitude.

This example illustrates a number of interesting features.  First, the
polynomial $\phi$ is of order two, as compared with the order three
polynomials that arose with the Hadamard-Toffoli gate set.  Second,
the way in which we do arithmetic is substantially more complex than
in the Hadamard-Toffoli case.  Third, we now have only to count
solutions to a single polynomial equation, instead of a set of
simultaneous polynomial equations.  There three features illustrate a
general fact: choosing different sets of universal gates gives rise to
sets of polynomial equations with different structures.  It is an
interesting problem to find gate sets giving rise to particularly nice
sets of equations in the sum-over-paths approach.

%
% improvements of our result
%
We mention one final variation on our result that seems worth
pursuing. This is a sum-over-paths approach to the Heisenberg
representation of quantum computation~\cite{Gottesman99b}, i.e.,
using the stabiliser formalism~\cite{Gottesman97a,Nielsen00a}. In
this formalism the $|\mathbf{0}\rangle$ state of $N$ qubits is
described as the simultaneous $+1$ eigenspace of the stabilizer
generators $\mathcal{S}_0=\langle Z_1,\ldots,Z_N\rangle$.
Subsequent evolution through a quantum circuit $U_x$ is described
by conjugating each of the stabilizer generators by the unitary
$U_x$.  It is well known\footnote{See, e.g., the discussion of the
Gottesman-Knill
  theorem in Chapter~10 of~\cite{Nielsen00a}.} that the effect of
Clifford group gates such as the Hadamard and {\sc cnot} is to take
products of Pauli matrices (like $Z_1,\ldots,Z_N$) to other products
of Pauli matrices under conjugation, and for a circuit entirely made
up of Clifford group gates this enables us to easily determine the
final state at the end of the computation.  However, the Clifford
group gates are not universal on their own.  To get a universal set,
we need to add another gate, such as the $T$ gate, which when acting
on $X$, $Y$, and $Z$ induces the following transformations:
\begin{equation}
\begin{split}
TXT^\dag = & \frac{X+Y}{\sqrt2}, \quad TYT^\dag = \frac{-X+Y}{\sqrt2},
\\ & \quad\text{and}\quad\quad TZT^\dag = Z.
\end{split}
\end{equation}
It is now possible to apply a sum-over-paths approach to each of the
stabilizer generators, with the $T$ gate playing a similar role to
that played by $H$ in the sum-over-paths approach based on quantum
states.  In this case each ``path'' is a sequence $\mathcal{S}_j$ of
sets of stabiliser generators.  A ``matrix element'' is the amplitude
to induce a transition between the initial set of (Pauli) stabiliser
generators $\mathcal{S}_0$ to some final set of (Pauli) stabiliser
generators $\mathcal{S}_{m-1}$. Thus determining the stabilizer
generators at the end of a computation is equivalent to counting the
number of solutions to certain sets of polynomial equations.  Of
course, even if that could be done, there would still remain the
difficulty of working out the measurement statistics resulting from a
measurement of the final state output at the end of the circuit.
Nonetheless, this alternative description may provide a different
insight into the complexity of quantum circuit simulation.

\section{Conclusion}
\label{sec:conclusion}

In conclusion, we have shown that quantum computation is intimately
connected with the problem of counting the number of solutions to sets
of polynomial equations, i.e., to counting points on an algebraic
variety.  This is a well-known problem in mathematics, and one of the
central problems of algebraic geometry; even for the case where there
is just a single polynomial, this problem is connected to deep topics
such as the Weil conjectures.  In the context of computational
complexity, it is known that the problem of counting solutions to
polynomials over finite fields is $\mathbf{\#P}$-complete.  It is
possible that better understanding the structure of the polynomial
equations associated with quantum computations may result in further
insight into the computational power of quantum computers.

\section*{Acknowledgements}

MAN thanks Richard Cleve, Manny Knill and Mike Mosca for encouraging
and informative discussions about the subject of this paper. MAN is
especially grateful to Richard Cleve for alerting him to the existence
of the work by Ehrenfeucht and Karpinski, and for discussions on the
universality of Hadamard and Toffoli.  TJO thanks Wim van Dam, Andreas
Winter, and Simone Severini for encouraging discussions.

%%\bibliographystyle{unsrt}
%%\bibliography{mybib}

\end{document}